# Understanding Protein-Complex Assembly through Grand Canonical Maximum Entropy Modeling

Andrei G. Gasic[1,2], Atrayee Sarkar[1,2], and Margaret S. Cheung[1,2,3]*

[1]*Department of Physics, University of Houston, Houston, Texas, 77204, United States. and*
[2]*Center for Theoretical Biological Physics, Rice University, Houston, Texas, 77005, United States.*
[3]*Pacific Northwest National Laboratory, Seattle Research Center, Seattle WA, United States*

Abstract

Inside a cell, heterotypic proteins assemble in inhomogeneous, crowded systems where the abundance of these proteins vary with cell types. While some protein complexes form putative structures that can be visualized with imaging, there are far more protein complexes that are yet to be solved because of their dynamic associations with one another. Yet, it is possible to infer these protein complexes through a physical model. However, it is often not clear to physicists what kind of data from biology is necessary for such a modeling endeavor. Here, we aim to model these clusters of coarse-grained protein assemblies from multiple subunits through the constraints of interactions among the subunits and the chemical potential of each subunit. We obtained the constraints on the interactions among subunits from the known protein structures. We inferred the chemical potential, that dictates the particle number distribution of each protein subunit, from the knowledge of protein abundance from experimental data. Guided by the maximum entropy principle, we formulate an inverse statistical mechanical method to infer the distribution of particle numbers from the data of protein abundance as chemical potentials for a grand canonical multi-component mixture. Using grand canonical Monte Carlo simulations, we captured a distribution of high-order clusters in a protein complex of Succinate Dehydrogenase (SDH) with four known subunits. The complexity of hierarchical clusters varies with the relative protein abundance of each subunit in distinctive cell types such as lung, heart, and brain. When the crowding content increases, we observed that crowding stabilizes emergent clusters that do not exist in dilute conditions. We, therefore, proposed a testable hypothesis that the hierarchical complexity of protein clusters on a molecular scale is a plausible biomarker of predicting the phenotypes of a cell.

*margaret.cheung@pnnl.gov

# I. INTRODUCTION

Living cells can contain on the order of $10^4$ (1) distinct types of proteins and other macromolecules at a given time. In this many-component mixture environment, macromolecules like proteins fold, unfold, and assemble into complexes and organize hierarchically into spatial networks(2-4). In fact, these unfathomably complex networks give rise to the emergence of all biological functions and ultimately the properties of life.(3, 5-7) The specific arrangements of macromolecules are thought to emerge from the vast amount of weak "quinary" and entropic interactions.(8-11) Of these types of interactions, the most intuitive conception of protein biophysics in this crowded environment is that of volume exclusion(12-14) exerted on a given protein by surrounding macromolecules, so called the "macromolecular crowding effects"(15). Proteins interact weakly and form higher-order complexes(16) through quinary interactions (17) where counter-acting forces between favorable electrostatic interactions and unfavorable solvation energies are provided by their metabolites.(18) Computer simulations have investigated the mechanism of these molecules forming small clusters often under the constraint of a fixed number of particles, *N*, in a closed system (i.e. a canonical ensemble in **Fig 1a**). This does not allow for particle fluctuations—a key free energy term. And yet, the physical mechanism of these complex assemblies and organization in an open system where *N* varies is still unclear.

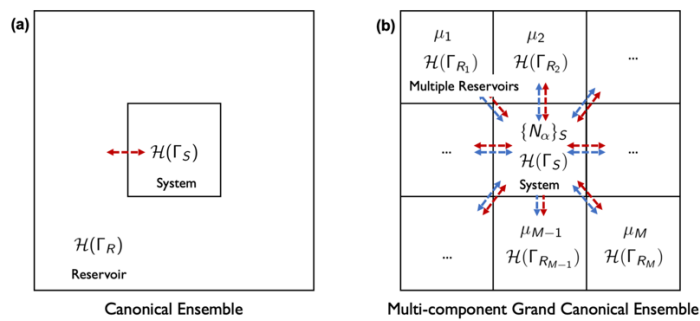

**FIG. 1**. (a) Canonical ensemble: the system *S* is maintained at a constant temperature through energy exchange (red arrow) with heat reservoir *R*. (b) Multi-component grand canonical ensemble: the system *S* is maintained at constant temperature and chemical potentials, $\{\mu_\alpha\}$, through energy (red arrow) and particle (blue arrows) exchanges with multiple particle reservoirs $R_\alpha$. $\mathcal{H}$ is the Hamiltonian with the set of positions and momenta, $\Gamma$, of the system or reservoir(s).

To resolve this issue, one may use semi-grand canonical ensembles (total $N$ is constant, but particle number of specific species fluctuates), which is especially important for studying phase separation of multi-component mixtures. However, there still remains the issue of setting the correct chemical potentials for each particle species that produces the correct stoichiometry or relative abundance(19, 20). Another way is to identify the constraint of mean particle numbers through a "chemical potential" in a grand canonical ensemble, thus allowing the $N$ to fluctuate (**Fig 1b**). However, it is challenging to establish constraints for chemical potentials for proteins in a multi-component mixture in an open system like cytoplasm. Here, we formulate a method to solve the inverse statistical mechanics problem of finding the correct chemical potentials for a multi-component mixture from the database of protein abundance. Current high-throughput experiments such as mass spectrometry quantify the protein abundance of a cell with high accuracy. It is often shown that the measurement of protein abundance is indicative of a cell's state(21-23).

There are several important reasons for using the grand canonical ensemble to study biological many-component systems instead of using the canonical ensemble:

- The assemblies of these macromolecular complexes are inherently finite-size processes where the abundance of a single protein species is experimentally measured in parts per million. Despite that a cellular mixture is highly crowded with biological molecules, the number of the individual species in this many-component mixture is still far from the thermodynamic limit. Without the correct ensemble, key free energy contributions will be missed(24, 25). Frequently, finite size corrections need to be used in canonical ensembles of protein binding for this reason(26, 27).
- In this investigation, we do not study phase transitions; however, our grand canonical method would be more advantageous in sampling various phases and mapping phase diagrams than an analogous canonical method. The grand canonical ensemble allows a single phase to occupy the entire simulation volume and avoiding the costly interfaces between phases(28-30). Also, with the grand canonical ensemble, it may reveal metastable free energy basins that are unstable phases in the corresponding canonical phase diagram(31).

- Information from the protein sequences provides a way of understanding protein-protein interaction networks(32, 33), and chemical potentials provides another source of rich information which dictates the state of a cell. While abundance may provide similar information, chemical potential may be more useful because the latter extends these systems into nonequilibrium or steady-state settings.

With the knowledge of protein abundance, we used grand canonical Monte Carlo simulations (GCMC) and the principle of maximum entropy to model the distinct features of protein assemblies. We use the mitochondrial respiratory Complex II or also called the Succinate Dehydrogenase (SDH) complex (**Fig. 2a**) to establish our model. It is a tetra-protein complex, composed of four subunits SDHA, SDHB, SDHC, and SDHD, and is well studied in a metabolic pathway that breaks down carbohydrates and produces energy.(34-37) Its integrative structure has also been constructed with constraints from the interactome through cross-linking mass spectrometry(38). SDH is an integral membrane protein complex, where SDHC and SDHD are inter-membrane proteins, and is in both the tricarboxylic acid cycle and aerobic respiration.(36, 37) To perform its various functions, the SDH form heterodimers and in some cases trimers before completely forming tetramer. Many other proteins and small molecules regulate the assembly and function of the SDH complex depending on the state or type of cell. As such, different cell types contain different abundances of each subunit.(21-23)

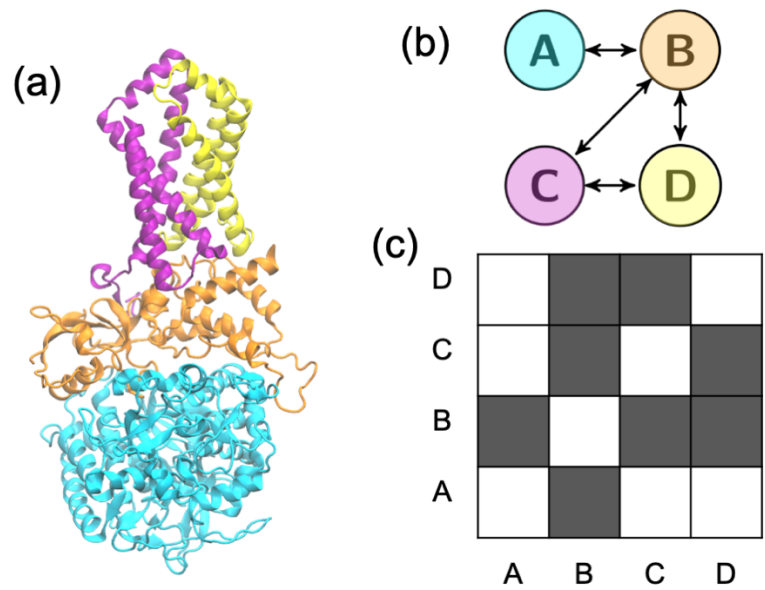

**FIG. 2.** (a) Crystal structure of Succinate Dehydrogenase (SDH) complex (protein data bank ID: 1NEN) with subunits SDHA through SDHD labelled A through D respectively. (b) Interaction topology of the SDH complex. Arrows indicate an interaction between the two subunits. (c) Contact map of the SDH complex, where white represents no interaction and grey represents an interaction between two of the subunits.

With the constraints of interaction topology among subunits and the constraints of abundance in the multi-component mixture, we are able to integrate this information into protein assemblies with tunable topological properties and expandable cluster sizes. By varying the crowding content, emergent clusters are formed where the transient complex exists at high crowding but not in lower crowding content. We have applied Jaccard index to characterize the higher order of these complexes and note that the emergent cluster for Lung and Brain is prominent and the representation of whole cell is not adequate. This integrative model shows how complex matter connects to the phenotype of a cell.

## II. THEORETICAL MODEL AND METHODS

### A. Abundance for Succinate Dehydrogenase (SDH)

The SDH complex from the PDB reveals four subunits interaction map are shown in **Fig. 2b and 2c**. The stoichiometry of each subunit is the same in an SDH complex(39); However, the abundance of each unit varies significantly among cell types(40). To understand how varying the abundances of the subunits affects the assembly of the complex, we study four different cell types in **Table I**. It shows these abundances in parts per million (ppm) for each of the subunits for whole (integrated average of all cell types over the whole organism), heart, brain, and lung cell types of mouse from the Protein Abundances Across Organisms Database (PaxDB)(41). **Table 1** also shows the relative abundance between the four subunits among these four cell types.

**TABEL 1.** Protein abundance $x_\alpha^{\text{exp}}$ in parts per million (ppm) and relative abundance $\tilde{x}_\alpha^{exp}$ (maximum of 1) for different mouse cell types from PaxDB(41).

| Type | Abundance $x_\alpha^{\text{exp}}$ (ppm) | | | | Relative abundance $\tilde{x}_\alpha^{\text{exp}}$ | | | |
|---|---|---|---|---|---|---|---|---|
| | A | B | C | D | A | B | C | D |
| Whole | 196 | 77.9 | 16.1 | 8.65 | 0.66 | 0.26 | 0.05 | 0.03 |
| Heart | 2358 | 637 | 108 | 2.88 | 0.76 | 0.21 | 0.03 | 0.001 |
| Brain | 853 | 87.2 | 56.7 | 2.9 | 0.85 | 0.09 | 0.06 | 0.003 |
| Lung | 495 | 2229 | 30.1 | 2.95 | 0.18 | 0.81 | 0.01 | 0.001 |

### B. Applying Principle of Maximum Entropy to the Grand Canonical Ensemble

Our approach is to infer a protein cluster model for the probability distribution of particle numbers over system states that are consistent with these experimental results with as little bias as possible. Using the principle of maximum entropy(42, 43) is a way to best guess the distribution which agrees with an average observable of the data. In our case, the probability distribution we seek is of the set of particle numbers $N_\alpha$ of species α, (or subunit), which is consistent with the experimentally derived protein abundance for α.

The distributions $P(\{N_\alpha\})$ is estimated by maximizing the entropy,

$$S[P] = -\sum_{\{N_\alpha\}} P(\{N_\alpha\}) \ln P(\{N_\alpha\}), \tag{1}$$

with constraints. This will give an exponential distribution for the energy landscape of the system that allows particle number fluctuations; i.e., a grand canonical distribution (see **Fig. 2b**).

$$P(\{N_\alpha\}) = \frac{Z(\{N_\alpha\})}{\Xi(\{\mu_\alpha\})} \exp\left(\frac{1}{k_B T} \sum_\alpha^M \mu_\alpha N_\alpha\right) \tag{2}$$

Here, the Lagrange multipliers for the constraints, i.e., the protein abundances, are chemical potentials $\mu_\alpha$ times inverse temperature ($\beta = 1/k_B T$). $Z(\{N_\alpha\})$ is the $N$-particle canonical partition function of a $M$-component protein mixture.

$$Z(\{N_\alpha\}) = \prod_i^N \int \exp\left(-\frac{\mathcal{H}_N(\{r_i\})}{k_B T}\right) dr_i \tag{3}$$

where, $\mathcal{H}_N$ is the $N$-particle Hamiltonian and $N = \sum_\alpha^M N_\alpha$. The distribution Eq. 2 is normalized by the grand canonical partition function,

$$\Xi(\mu) = \sum_{N_1=0}^\infty \cdots \sum_{N_M=0}^\infty \left(\prod_\alpha^M e^{\mu_\alpha N_\alpha/k_B T}\right) Z(\{N_\alpha\}). \tag{4}$$

Additionally, this model does not require parameter tuning; all parameters are completely determined by the experimental data.

### C. Parametrizing Chemical Potential Using Max Entropy

There exists a unique set of chemical potential $\{\mu_\alpha\}$ that produces observable mean particle numbers $\langle N_\alpha \rangle$ that is consistent with the experimentally measured $\langle N_\alpha \rangle^{\exp}$ (from the protein abundance database PaxDB at https://pax-db.org/ (41)), but finding them is a computationally difficult inverse statistical mechanics problem.

In order to solve for the correct chemical potential $\mu_\alpha$ for each particle type α, we minimize the Kullback-Leibler (KL) divergence between the "real" distribution, $Q$, and the model distribution (Eq. 2), defined as:

$$D_{KL}(Q||P) = \sum_{\{N_\alpha\}} Q(\{N_\alpha\}) \ln \frac{Q(\{N_\alpha\})}{P(\{N_\alpha\})} \tag{5}$$

The real distribution is the Boltzmann-like distribution containing $\langle N_\alpha \rangle^{\exp}$. The KL divergence becomes,

$$D_{KL}(Q \,||\, P) = \ln\frac{\Xi(\boldsymbol{\mu})}{\Xi_0} - \frac{1}{k_B T}\sum_\alpha^M \mu_\alpha \langle N_\alpha \rangle^{\exp} \tag{6}$$

where $\Xi_0 \equiv \Xi(\mu = 0)$ and $\boldsymbol{\mu} \equiv \{\mu_\alpha\}$.

The partition function ratio in the expression can be further simplified using a cumulant expansion as

$$\ln\frac{\Xi(\boldsymbol{\mu})}{\Xi_0} = \ln\frac{\sum_{\{N_\alpha\}} \prod_i^N \int e^{-\frac{\mathcal{H}_N}{k_B T}} e^{\frac{\sum_\alpha^M \mu_\alpha N_\alpha}{k_B T}} d\boldsymbol{r}_i}{\sum_{\{N_\alpha\}} \prod_i^N \int e^{-\frac{\mathcal{H}_N}{k_B T}} d\boldsymbol{r}_i} \tag{7}$$

$$= \ln\left\langle e^{\frac{\sum_\alpha^M \mu_\alpha N_\alpha}{k_B T}} \right\rangle_0 \tag{8}$$

$$= \sum_{\nu=1}^{\infty} \frac{1}{\nu!} \left\langle \left(\frac{1}{k_B T}\sum_\alpha^M \mu_\alpha N_\alpha\right)^\nu \right\rangle_c \tag{9}$$

Where $\langle \ldots \rangle_c$ signifies the cumulant. To second order, the cumulant expansion is

$$\ln\frac{\Xi(\boldsymbol{\mu})}{\Xi_0} = \frac{1}{k_B T}\sum_\alpha \mu_\alpha \langle N_\alpha \rangle \tag{10}$$
$$+ \frac{1}{2(k_B T)^2} \sum_{\alpha\beta} \mu_\alpha \mu_\beta \left(\langle N_\alpha N_\beta\rangle - \langle N_\alpha\rangle\langle N_\beta\rangle\right)$$

Plugging this back into the KL divergence in Eq. 6,

$$D_{KL} = \frac{1}{k_B T} \boldsymbol{\mu} \cdot (\langle \boldsymbol{N}\rangle - \langle \boldsymbol{N}\rangle^{\exp}) + \frac{1}{2(k_B T)^2} \boldsymbol{\mu}^T \cdot C \cdot \boldsymbol{\mu} \tag{11}$$

where $C_{\alpha\beta} = \langle N_\alpha N_\beta\rangle - \langle N_\alpha\rangle\langle N_\beta\rangle$. The solution for $\mu$ that minimizes $D_{KL}$ (i.e., $\frac{\delta D_{KL}}{\delta \mu} = 0$), is

$$\boldsymbol{\mu} = -k_B T \, C^{-1} \cdot (\langle \boldsymbol{N}\rangle - \langle \boldsymbol{N}\rangle^{\exp}) \tag{12}$$

Since this is an approximate solution, when $\langle N_\alpha \rangle \neq \langle N_\alpha \rangle^{\exp}$, we use the following iterative self-consistent procedure to find more accurate values that eventually would converge to reproduce the experimental measurements. The algorithm is as follows in Algorithm 1.

**Algorithm 1** Self-consistent algorithm

1: Set $\langle N_\alpha \rangle^{\exp}$ and $\mathcal{H}_N$ for system
2: Initialize: performing simulations with $\{\mu_\alpha\} = 0$
3: Estimate $\langle N_\alpha \rangle$ and $C_{\alpha\beta}$
4: **while** (error $= \sum_\alpha |\langle N_\alpha \rangle - \langle N_\alpha \rangle^{\exp}| / \sum_\alpha \langle N_\alpha \rangle^{\exp}$) > tolerance value, **do**
5:     Update: $\mu \to \mu - \frac{1}{\beta} C^{-1} \cdot (\langle \mathbf{N} \rangle - \langle \mathbf{N} \rangle^{\exp})$
6:     Repeat simulations with updated $\{\mu_\alpha\}$
7:     Estimate new $\langle N_\alpha \rangle$ and $C_{\alpha\beta}$
8: **end while**

Additionally, since PaxDB records protein abundance, $x_\alpha^{\exp}$, in units of part per million (ppm) and our simulation box size is a small fraction of the size of a cell, we determine $\langle N_\alpha \rangle^{\exp}$ from the relative abundance for a certain volume fraction $\phi$ ($\equiv \frac{Nv}{V_{\text{box}}}$, where $v$ is the volume of the protein and $V_{\text{box}}$ is the volume of the box). That is,

$$\langle N_\alpha \rangle^{\exp} \equiv \frac{x_\alpha^{\exp}}{\sum_\alpha^M x_\alpha^{\exp}} \frac{\phi V_{\text{box}}}{v}. \tag{13}$$

### D. Grand Canonical Hamiltonian

To model protein assemblies *in vivo*, we will join aspects of both models used in chromosomes(44, 45) and molecule self-assembly modeling(46). We used structure-based Hamiltonian, i.e., the model has attractive interactions between proteins that are in contact in the crystal structure (**Fig. 2a**), and there are volume exclusion interactions for proteins not in contact. The $N$-particle Hamiltonian is a function of the particle positions $\{r_i^\alpha\}$ of particle species α, having the following form:

$$\mathcal{H}_N(\{\mathbf{r}_i^\alpha\}) = \sum_{\alpha<\beta}^{M} \sum_i^{N_\alpha} \sum_j^{N_\beta} U_{\text{LJ}}(\mathbf{r}_i^\alpha, \mathbf{r}_j^\beta) \Theta(\Delta_{\alpha\beta} - |\mathbf{r}_i^\alpha - \mathbf{r}_j^\beta|) \tag{14}$$

where our effective pairwise protein potential is the Lennard-Jones potential $U_{\text{LJ}}$,

$$U_{\text{LJ}}\left(\boldsymbol{r}_i^\alpha, \boldsymbol{r}_j^\beta\right) = 4\epsilon \left[ \left( \frac{\sigma_{\alpha\beta}}{\left|\boldsymbol{r}_i^\alpha - \boldsymbol{r}_j^\beta\right|} \right)^{12} - \left( \frac{\sigma_{\alpha\beta}}{\left|\boldsymbol{r}_i^\alpha - \boldsymbol{r}_j^\beta\right|} \right)^{6} \right] \quad (15)$$

For simplicity, $\sigma_{\alpha\beta} = \sigma$. The term $\Delta_{\alpha\beta}$ in the Heaviside function $\Theta$ is the cutoff value between particle species $\alpha$ and $\beta$. Since our contact data is binary (probability of 1 or 0 of being connected), we use

$$\Delta_{\alpha\beta} = \begin{cases} 2.5\sigma & \text{if } \alpha, \beta \text{ interact specifically} \\ 1\sigma & \text{otherwise} \end{cases} \quad (16)$$

The interaction strength $\epsilon = 0.5$ is used to ensure the stability of the system. To ensure a one-to-one correspondence of $\mu$ to $\phi$, we need to select the best interaction strength $\epsilon$. For $\epsilon \gtrsim 0.8$, certain values of $\mu$ correspond to multiple $\phi$ values (see Fig A1.), which signify multiple phases. Thus, a one-to-one correspondence of $\mu$ to $\phi$ reduces the chance of having large fluctuation in energy and density from phase transitions.

### E. Monte Carlo Simulation Details

We conducted grand canonical Monte Carlo (gcmc) simulations on LAMMPS(47) using a box size $V_{\text{box}} = 1000\sigma^3$ for 120,000 gcmc steps, where each gcmc step attempted 800 insertions or deletions and 800 translations per particle type. The mean energy and variance of particle number plateaus at approximately 20,000 steps, and the errors for the mean energy and variance of particle number plateaus at approximately 100,000 steps. Thus, 120,000 steps are sufficient for data analysis. The number for insertions or deletions and translations ensures that the autocorrelation time for energy and density is one gcmc step. The attempts in the gcmc steps follow the metropolis criterion,

$$p(\mathcal{H}_N \to \mathcal{H}'_N) = \min\{1, \exp[-(\mathcal{H}'_N - \mathcal{H}_N)/k_B T]\}$$

for translation, and

$$p(N \to N') = \begin{cases} \min\left\{1, \dfrac{V}{(N+1)} \exp[-(\mathcal{H}_{N+1} - \mathcal{H}_N - \mu)/k_B T]\right\} & \text{for } N' = N+1 \\ \min\left\{1, \dfrac{N}{V} \exp[-(\mathcal{H}_{N-1} - \mathcal{H}_N + \mu)/k_B T]\right\} & \text{for } N' = N-1 \end{cases}$$

for insertion and deletion of particles.

### F. Jaccard Index

We followed the analysis by Sardiu *et al* (48, 49) who have used the Jaccard index, J, to analyze and predict core components and modules in higher-order protein complexes. The Jaccard index, J, measures similarity among sample sets. J is defined as the size of the intersection divided by the size of the union of the sample sets. In our case we have four sample sets corresponding to the four subunits A, B, C and D. Each set contains all the subunit pairs interacting with the subunit corresponding to that set. For instance, the sample set corresponding to A contains subunits pairs AA, AB, AC, and AD while the set corresponding to B contains AB, BB, BC, and BD. We follow the number of contacts formed by a pair of subunit α and subunit β, $n_{\alpha\beta}$, among the four subunits, A, B, C, and D. A pair of subunits α and β are considered to be interacting (i.e., forming a pair) if their Euclidean distance is less than a predefined cut off value of 1.1 σ. The Jaccard index J(α,β) between subunits α and β is below

$$J(\alpha, \beta) = \frac{n_{\alpha\beta}}{\sum_\gamma (n_{\alpha\gamma} + n_{\beta\gamma}) - n_{\alpha\beta}}$$

We then calculated the time average of ⟨J(α,β)⟩ for each cell types.

## III. RESULTS

### A. A self-consistent algorithm converges to chemical potentials according to mean abundances of multi-component protein-complex

In accordance with the principle of maximum entropy (discussed in Section IIB), we inferred the set of chemical potentials for each of the subunits of the SDH complex, $\{\mu_\alpha\}$ (where α ≡ A, B, C, D corresponding to SDHA, SDHB, SDHC, and SDHD, respectively), such that the mean particle numbers $\langle N_\alpha \rangle$ matches with the relative abundance from PaxDB in **Table I**. We used our self-consistent algorithm (**Algorithm 1**) to compute $\{\mu_\alpha\}$ for each cell type at three different volume fractions, φ = 0.1, 0.2, and 0.3. These three volume fractions represent the range in the fraction of macromolecular volumes in a cellular volume where macromolecular crowding(15) affects protein dynamics(13, 50) and assemblies. On average, the algorithm converges to $\{\mu_\alpha\}$ of each system in under three iterations.

As a demonstration of our self-consistent algorithm (**Algorithm 1**), we present the evolution of the sets $\{\mu_\alpha\}$ and $\{\langle N_\alpha \rangle\}$ for the Whole cell type at a volume fraction $\varphi = 0.1$. **Fig.3(a)** shows $\{\mu_\alpha\}$ changing with each iteration of the algorithm resulting in a $\langle N_\alpha \rangle$ [**Fig.3(b)**]. Starting with $\{\mu_\alpha\} = 0$, the error [**Fig.3(a)** right y-axis] is large due to the large deviations of $\langle N_\alpha \rangle$ from $\langle N \rangle^{\text{exp}}$ [**Fig.3(b)**]. By the third iteration of **Algorithm 1**, $\{\mu_\alpha\}$ results in a simulation that captures the correct experimental particle means $\langle N \rangle^{\text{exp}}$ values for each subunit.

This algorithm converges to $\{\mu_\alpha\}$ of the different cell types (**Table I**) and different volume fractions in a similar fashion. The values of the converged $\{\mu_\alpha\}$ are shown in the next section.

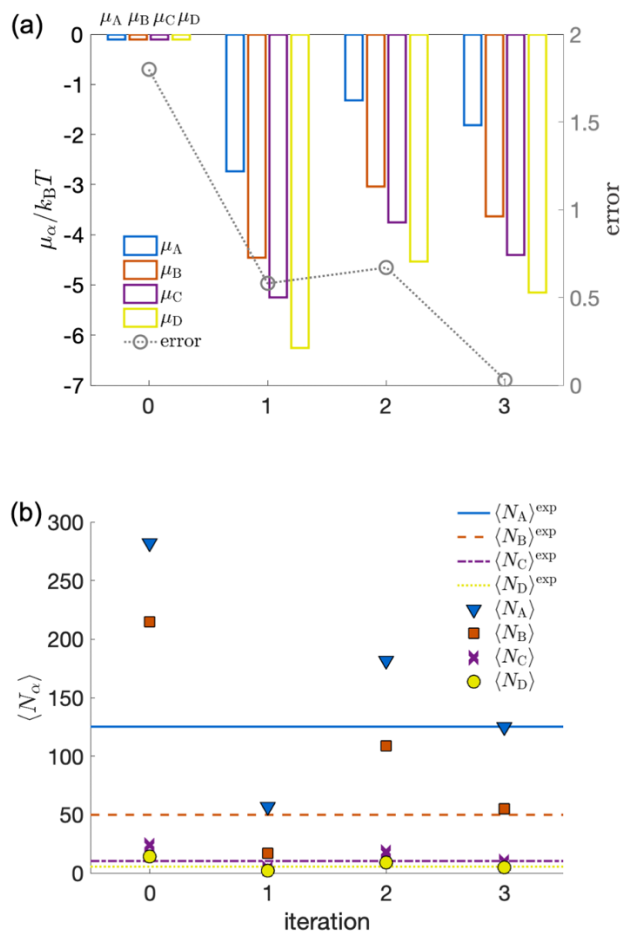

**FIG 3.** (a) Values of $\mu_\alpha$ and (b) $\langle N_\alpha \rangle$ where $\alpha \equiv$ A, B, C, D corresponding to subunits SDHA, SDHB, SDHC, and SDHD, respectively, vs iterations of the self-consistent algorithm for SDH complex of whole cell type at $\phi=0.1$. Dashed lines indicate experimentally observed average particle numbers. (a- right axis) Error vs iterations in gray.

## B. Cell Type is Distinguished by Chemical Potentials of Subunits

Since each cell type has varying protein abundances for each subunit of the SDH complex (as shown in **Table I**), we hypothesized that the chemical potential will change to capture the correct statistics accordingly. To show the effects on the μα values when cell type varies, we calculate the μα using **Algorithm 1** for cell types Whole, Lung, Heart, and Brain at $\phi$ = 0.1, 0.2, and 0.3. The converged $\mu_\alpha$ values are shown in **Fig.4** (presented as symbols) as a function of $\phi$. The solid line curves are analytical approximations (see **Appendix A**) of Eq. A3. The μα values calculated via **Algorithm 1** match well with the analytical results of Eq. A3. This agreement confirms that **Algorithm 1** works correctly. The slight deviations the algorithm computed values from Eq. A3 may be due to the non-zero error of **Algorithm 1** or that Eq. A3 is not an exact result.

Indeed, each cell type gives a unique set $\{\mu_\alpha\}$ shown in **Fig. 4**. These chemical potentials are the subunit free energy, resulting in a unique landscape for each cell type. Since subunit B interacts with all the other subunits (see **Fig. 2**), $\mu_B(\phi)$ will behave differently from the others. The general trends are that $\mu_B$ is flat for all values of $\phi$, while the chemical potentials of the other subunits are monotonically increasing.

The cell type with chemical potential trends that is the most strikingly different from others are that of the Lung cell type in **Fig.4(c).** In the Lung cell type, the $\mu_B$ curve (orange), it monotonically increasing unlike that of the other cell types. Such effect is due to the large relative abundance of SDHB ($\tilde{x}_B^{exp}$) dominating in the Lung cell type by a factor of 4.5 times the next largest relative abundance, which is SDHA. In the other cases, $\tilde{x}_A^{exp}$ is at least 2.5 times larger than $\tilde{x}_B^{exp}$ (see **Table I** for exact numbers).

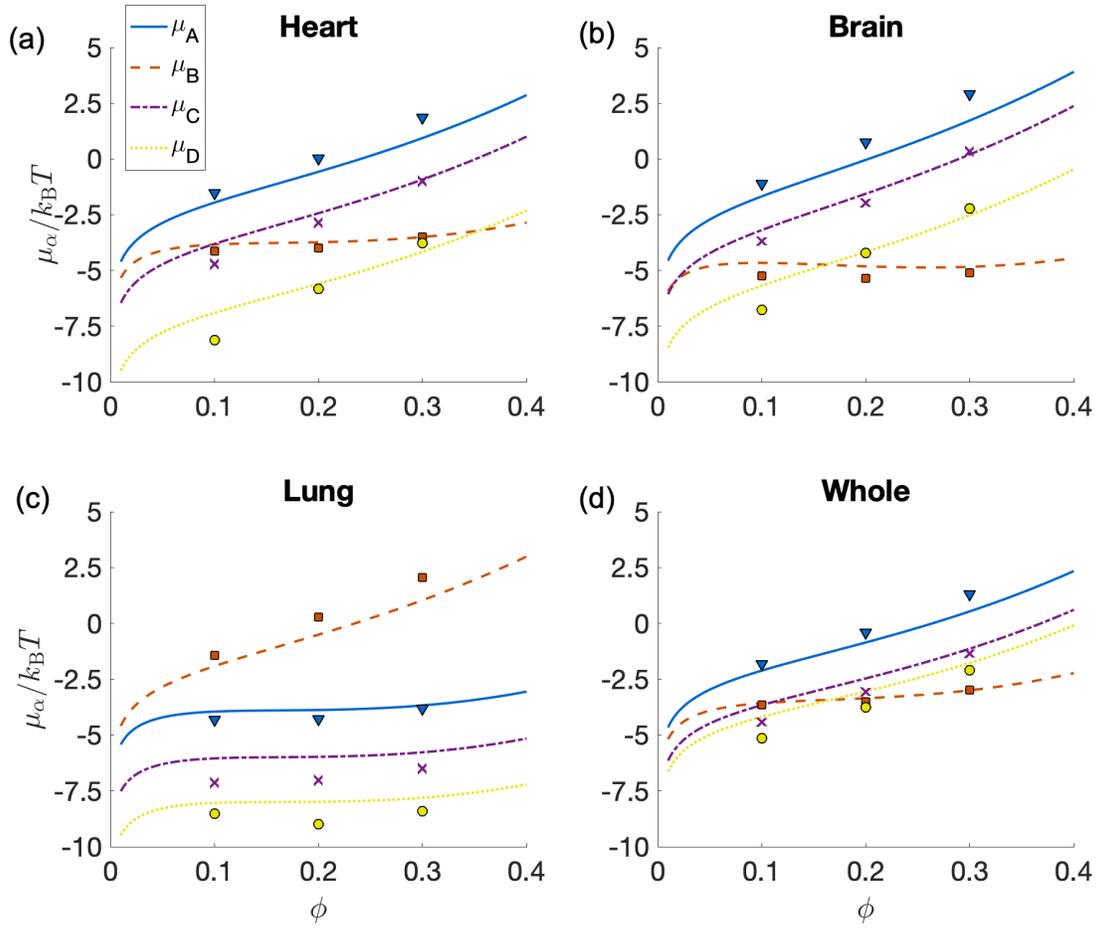

**FIG 4**. Chemical potential as a function of total volume fraction (ϕ) using for various cell types: (a) Heart, (b) Brain, (c) Lung, and (d) Whole. Solid lines are the analytical approximations from Eq. A3, and symbols (triangle: A, square: B, cross: C, and circle: D) are values of $\mu$ from calculated from the self-consistent algorithm (**Algorithm 1**).

### C. The radial distribution of contact hub

As the results from the previous section showed that $\mu_B(\phi)$ behaves differently from the other subunits, we focus our attention to subunit B. To further understand the effect of changing cell types, we extend our analysis to calculating the radial distribution g(r) between subunit pairs B and itself ($g_{B-B}$), B and the other non-B subunits ($g_{B-!B}$), and all pairs ($g_{tot}$) for ϕ=0.1 and 0.3, shown in Fig. 5 and 6, respectively.

At $\phi = 0.1$ (**Fig. 5**), the radial distribution function of inter particle distance, *r*, between any two pairs of subunits, $g_{tot}(r)$ resembles the that of a gas or dilute liquid for all four cell types. Its first peak, at $r \approx 1.2$ σ, is approximately 1.5 times the average density of the system. The change in the first peak of the curves can be thought as a change in an "osmotic pressure" between two particle types, which is dictated by both entropic and energetic effects. Since B is the contact hub (all subunits interact with B; see **Fig. 2**), the first peak of $g_{B-!B}(r)$ is approximately 70% larger in amplitude than that of $g_{tot}(r)$) regardless of the cell type. This first peak amplitude increase is due to the preferred interaction with subunit B than with any other random particle of the multi-component mixture. Whereas, the first peak of the $g_{B-B}(r)$ curve is only slightly less than that of $g_{tot}(r)$, resembling a hard sphere gas.

At $\phi$ =0.3 (**Fig. 6**), the radial distribution function plots have emerged more pronounced peaks than those curves at $\phi$ =0.1 in **Fig 5**. Here, a distinct second peak appears, characterizing a higher-order structure. The first peak of $g_{B-!B}(r)$ is approximately 25% larger in amplitude than that of $g_{tot}(r)$ regardless of the cell type. Again, this first peak amplitude increase is due to the preferred interaction with subunit B. However, the amplitude increases of the first peak of $g_{B-!B}(r)$ is less than that compared to $g_{tot}(r)$ at $\phi$ =0.1. This outcome is due to the more prominent entropic forces at $\phi$ =0.3 than at $\phi$ =0.1.

Interestingly, unlike at $\phi$ =0.1, at $\phi$ =0.3, the $g_{B-B}(r)$ curve differs between the cell types. For instance, the effect of changing the cell type either increases the osmotic pressure for Lung in **Fig.6(c)** or decreases it for Brain in **Fig.6(b)**. The highest first peak value for $g_{B-B}(r)$ is that of the Lungs. B is closer to itself than in the other cell types due to the relative increase in $\mu_B$ in **Fig. 4**. Even though subunit B does have energetic self-interactions other than volume exclusion, the growth in $\mu_B$ increases the entropic forces (or osmotic pressure) between itself due to macromolecular crowding. The lowest first peak value of $g_{B-B}(r)$ is that of the Brain, signifying a decrease in entropic interaction. Interestingly, the second peak of $g_{B-B}(r)$ for the Brain is the largest in amplitude, signifying that the B subunits are assembling together with another subunit type in between. This points toward subunits forming *emergent* higher-order assemblies under high macromolecular crowding content.

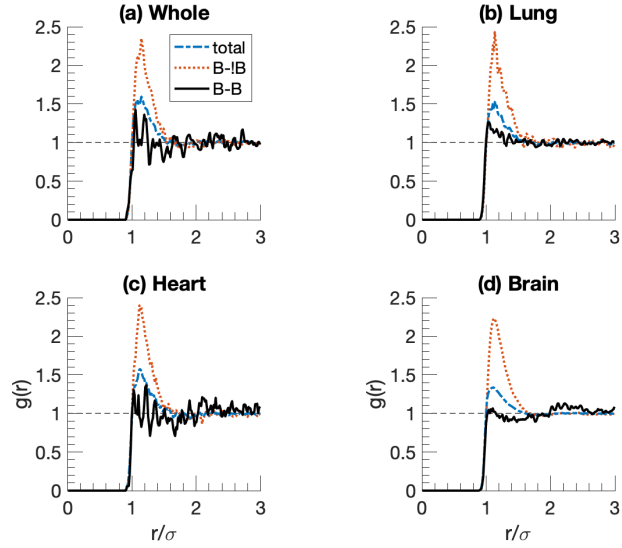

FIG 5. Radial distribution function $g(r)$ between all particles (blue dash-dot, total), B and not B (orange dotted, B-!B), and B and itself (black solid line, B-B) for various cell types at $\phi = 0.1$: (a) Lung, (b) Heart, (c) Brain, and (d) Whole. The graphs are smoothed using running averages using mean over a window of 3-time frames.

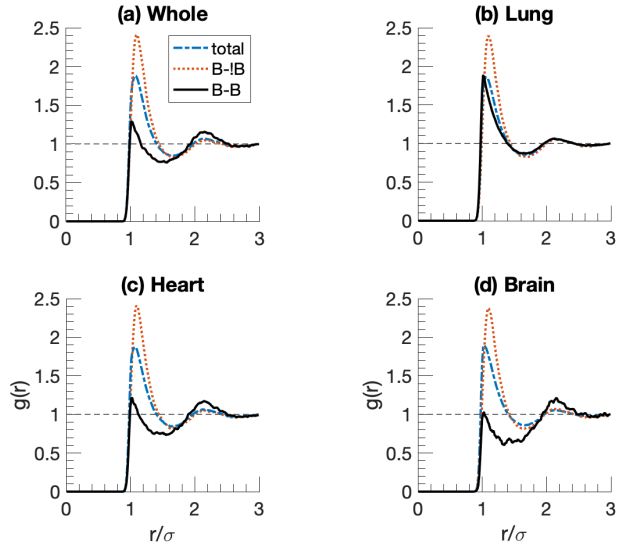

FIG 6. Radial distribution function $g(r)$ between all particles (blue dash-dot line, total), B and not B (orange dotted line, B-!B), and B and itself (black solid line, B-B) for various cell types at $\phi = 0.3$: (a) Lung, (b) Heart, (c) Brain, and (d) Whole. The graphs are smoothed using running averages using mean over a window of 3 time frames.

### D. Higher Order Assemblies Built from Pair-wise Information

Because of crowding and the certain preferential interactions between subunits, the mixture is highly inhomogeneous, filling with particles in lumps and clusters. The variation in the profile in $g(r)$ signifies the presence of emergent clustering of the subunits around B at high crowding content at $\phi$ =0.3, which does not exist at low crowding content at $\phi$ =0.1. Next, we gained more insight into the higher-order structure of the assemblies by calculating the Jaccard index, J, using Eq. 19. The use of J is a well-established computational-topology algorithm to identify a new protein family or a gene from existing database by comparing similarities in their interacting networks(51). Sardiu *et al* have applied this index to determine whether a protein belongs to a core of protein networks in an interactome database as a justification of forming physical assemblies(48, 49). Here, we leveraged the Jaccard index to further measure whether a subunit belongs to an emergent cluster of heterotypic particles. If $J(\alpha, \beta)$ = 1, the subunits $\alpha$ and $\beta$ only associate with each other and no other subunit. If $J(\alpha, \beta)$ = 0, the subunits $\alpha$ and $\beta$ only associate with other subunits or they are not found in the system for the particular time. Thus, Eq. 19 is the probability of subunits $\alpha$ and $\beta$ associating with each other given that the two subunits are not isolated and part of a complex assembly. Comparing this with the contact matrix in **Fig. 2(c)**, we can identify the emergent properties of forming a cluster that stems from the chemical potentials, or macromolecular crowding, instead of specific protein-protein interactions. Since the heart cell type has a similar $g(r)$ profile as the whole cell type, we focused this analysis on whole, brain, and lung cell types.

First, we examine the time average $<J(\alpha, \beta)>$ at a low crowding content at $\phi$ =0.1 in **Fig.7(a-c)**. The J index highest values are the A-B pair in Whole; whereas in Lung and Brain, it is in the self-interaction B-B and A-A, respectively. Since there are no self-interactions of the subunits on the contact matrix [**Fig. 2(c)**], these highest J values in Lung and Brain are purely driven by the relative abundance and hence chemical potentials. A representative snapshot with the largest cluster highlight is shown **in Fig.7(d-f)**. In these plots, A, B, C and D subunit types are color coded as cyan, orange, purple and yellow, respectively. Furthermore, in comparison with the Whole cell type [Figs. 7(a)], the Lung and Brain J matrices do not average to that of Whole.

The clusters formed for the Whole cell type resemble the structure in **Fig. 2(b)**, but the Lung and Brain have clusters of mainly B and A subunits, respectively.

Beyond the diagonal of the J matrices, we also see nontrivial off-diagonal variations in J values between the cell types. The largest off-diagonal values for all cell types is the J(A,B), which can be attributed to the fact that subunits A and B have the two largest abundances (**Table 1**) and chemical potentials (**Fig. 4**) in all cell types. A key difference, though, between Lung and Brain is the increase in the J(A,C), J(B,C), and J(C,C) values for both $\phi$ =0.1 [**Figs. 7(b & c)**].

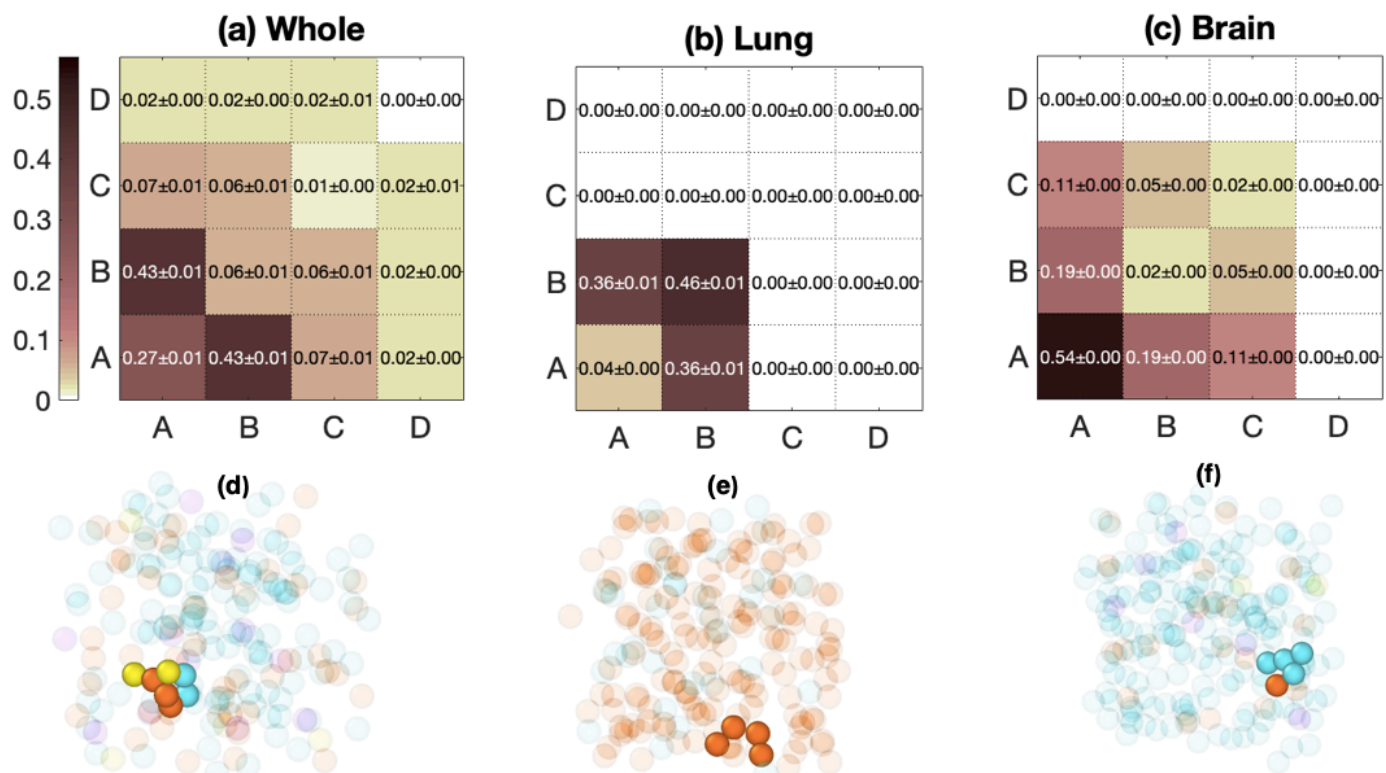

FIG 7. J matrix for (a) Whole, (b) Lung and (c) Brain cell types for $\phi = 0.1$. Panels (d), (e), and (f) show snapshots from randomly selected time frames of the clusters for respective cell types. The largest clusters have been highlighted. Subunit types A, B, C and D are color coded as cyan, orange, purple and yellow respectively. Contact cutoff is $1.1\sigma$, and visualization of the system is obtained using OVITO(52).

Next, we increase the crowding content from $\phi = 0.1$ to $\phi = 0.3$. Similar trends on J index persist at $\phi = 0.3$ for all cell types [**Fig. 8(a-c)**]. However, the higher volume fraction gives rise to emergent higher-order assemblies seen in **Fig. 8(d-f)**. Again in Lung, J(B,B) has the maximal value [**Fig. 8(b)**], and J(A,A) has the maximal value [**Fig. 8 (c)**] in Brain. Since entropic forces are heightened by increasing N, at $\phi = 0.3$ [**Figs. 8 (b & c)**], the maximum J value has increased to 0.51 for Lung and 0.57 for Brain.

Lastly, comparing the J values between $\phi = 0.1$ and $\phi = 0.3$, the general shifts in all cell types are toward the subunit pairs in which have no attractive interaction (the pairs in white in **Fig. 2c**) in the Hamiltonian (Eq. 14-15) and away from the subunit pairs that do (the pairs in gray in **Fig. 2c**). The increase in entropic interactions (i.e. depletion forces from crowding), due to the increase in N, causes this shift. Even though the main changes in relative abundance between the cell types are of subunits A and B (**Table 1**), the entropic interactions affect the J values of all the subunits nontrivially. At $\phi = 0.3$, higher-order clusters emerge with distinct features shown by the representative snapshots in Fig. **8(d-f).** In Lung, the dominant cluster with transient stability is composed of subunit B (**Fig 8e**). As crowding content reduces, such a cluster breaks down (**Fig 7e**). Such features permit the development of hypotheses connecting the properties of molecular assemblies to cell phenotypes.

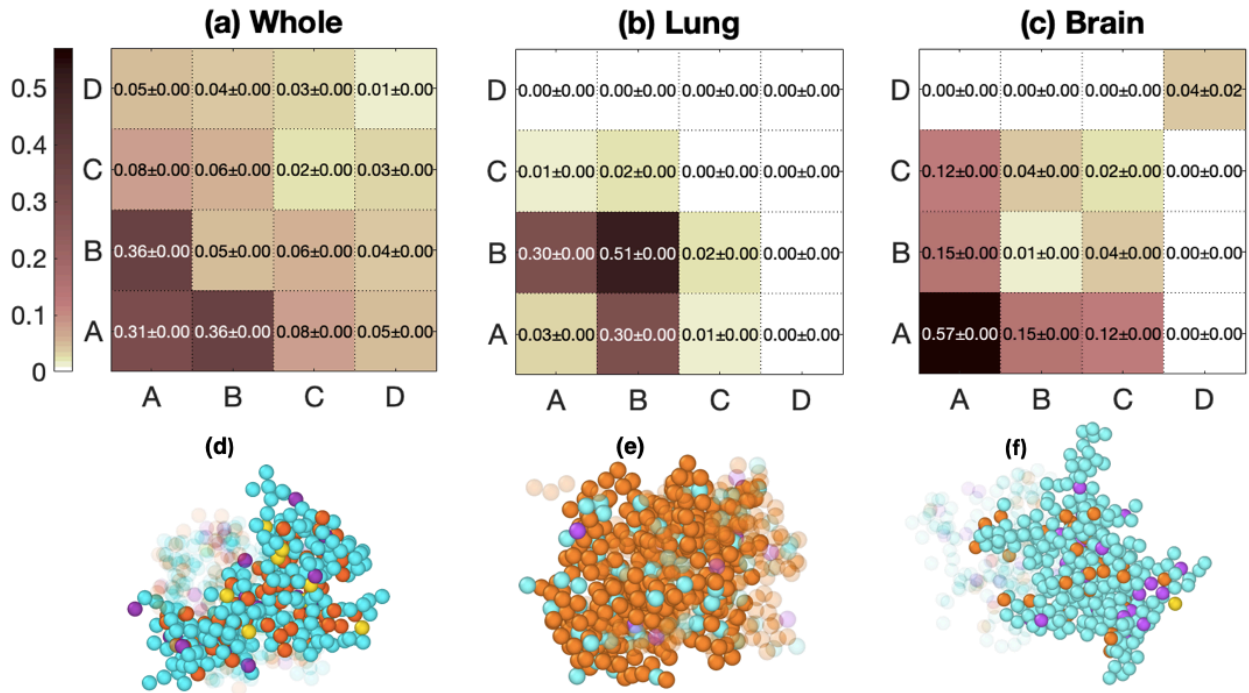

FIG 8. J matrix for (a) Whole, (b) Lung and (c) Brain cell types for $\phi = 0.3$. Panels (d), (e), and (f) show snapshots from randomly selected time frames of the clusters for respective cell types. The largest clusters have been highlighted. Subunit types A, B, C and D are color-coded as cyan,

orange, purple and yellow respectively. Contact cutoff is 1.1 $\sigma$, and visualization of the system is obtained using OVITO(52).

I. DISCUSSION

A. Leverage protein abundance as a constraint on the particle numbers by combining Grand Canonical Monte Carlo simulations with Max Entropy Inference

Protein assembly in the cell is thermodynamically governed by various enthalpic and entropic factors. In this investigation, we have allowed the particle number to fluctuate (i.e. the grand canonical ensemble; see **Fig 1**.) as an additional free energy contribution to understand the effects on protein assembly. The difficulty in using the grand canonical ensemble is that the chemical potentials for specific particle types are unknown. To avoid this problem, previous efforts in studying protein or synthetic particle assembly have either defined the chemical potentials as known(29-31, 46, 53, 54) or used semi-grand canonical ensembles(55), where particle types may vary in particle number with the total $N$ is fixed.

To our knowledge this is the first study to leverage the protein abundance attained from PaxDb;(41) as a constraint to solve for the chemical potential of a protein type in an open, crowded environment with the tools of inverse statistical mechanics(56), grand canonical Monte Carlo simulations, the maximum entropy principle(42, 43). The grand canonical ensemble allows particle numbers to fluctuate by placing particle reservoirs with fixed chemical potentials of each particle type in contact with the multi-component system (**Fig 1**).

By using the grand canonical ensemble, the fixed particle number is switched for a fixed chemical potential, trading one unknown for another. Analytically calculating the chemical potential is possible for simple systems such as hard spheres; however, this becomes increasingly difficult or impossible with varying protein-protein interactions, polydisperse mixture, or flexible polymers. To our knowledge, this approach is the only method that will approximate the correct chemical potentials given the particle number distribution. Our self-consistent algorithm (**Algorithm 1**) has proven to be a useful way to calculate the chemical potentials of the particles in a multi-component mixture from protein abundance. In principle, our method may be used for more complex systems such as a mixture differing particle shapes or macromolecules (or polymers) instead of simple spheres with the same radius.

### B. GCMC allows for the investigation of emergent complex formation in crowded media.

Here, in our study, the use of the Grand Canonical ensemble allows variation in particle number which in turn allows the possibility of having different chemical potentials for different cell types. This difference in chemical potential and variation in the number of particles in the simulation leads to the formation of emergent, higher-order complex structures. This feature is different from the simulation approaches based on a conventional canonical ensemble where the total particle number in a simulation box is fixed(19, 20). Previous inverse design studies have learned interaction potentials using radial distribution function $g(r)$ in a canonical ensemble. The interactions of the protein complex in a canonical ensemble become the main driving force that guides the formation of the complex structure.

By increasing the volume fraction of the system, the entropically favored assembly increases as well. Increasing the volume fraction (or crowding) of the system and keeping other physical properties constant will only change the entropy, since the crowding effect or depletion force is an entropic effect(12, 13, 57). In doing so, specific protein-protein entropic forces or "osmotic pressures" are created and varied by the level of crowding(13, 50). This osmotic pressure can be seen with the radial distribution function $g(r)$ since the integral of $g(r)$ gives the corrections to the ideal gas pressure in **Figs. 5 & 6**, dictated by a combination of entropic forces such as macromolecular crowding and specific interactions. Interestingly, even at low crowding content (**Fig 5**) the $g(r)$ between subunits B and not B subunits (i.e. B and !B) shows deviations from the total system regardless of the cell type by comparing to the contact map in **Fig 2**, which signifies specific protein-protein osmotic pressure that is controlled by the chemical potentials in a crowded environment where particle numbers are allowed to fluctuate.

Because of differences in abundance and interactions among subunits, the mixture is highly inhomogeneous with small lumps and clusters. This emergence of new cluster formation in a grand canonical ensemble is heightened when the crowding content increases to $\phi=0.3$ in **Fig. 8(d-f)** from 0.1 in **Fig. 7(d-f)**. We show that particularly for the Lung where the abundance of subunit B is much higher than those for other cell types, the emergent complex with lumps of B subunits is most significant at $\phi=0.3$, while other cell types do not show prominent lumps of B subunits.

### C. **Our investigations allow developing hypotheses connecting the high-order protein complexes with the cell phenotype**

Cell types may have the same gene expression, but the cell state will differ in protein abundance. (58) As the mixture with heterotypic particles is inhomogeneous, it is challenging to characterize the lumps or clusters in the system as densities of heterotypic particles alone is not sufficient to address their association with one another in space. To our knowledge, we are the first to use the Jaccard index to elucidate the higher-order complex structures from particle interactions. This measurement is key to uncover the topology of these complex structures, reflecting the differences in cell type.

From the J matrices (**Figs. 7 & 8**), we have seen that the Whole cell type does not resemble the specialized cell types (Lung and Brain). These distinct behaviors between the cell types may be the reason why proteins that are similar in structure and sequence form different complexes. For example, SDH forms trimers of the complex in E. coli,(34) but is only a single monomer in the porcine heart.(35) Our method may be used to study the various cell phenotypes leading from the higher-order complex assembly that is dictated from protein abundance, or the chemical potentials, of the protein species. The unique state or phenotype is connected to the unique chemical potential landscape (**Fig. 4**), which gives rise to emergent molecular topologies of higher-order complexes, depending on the crowding content.

## II. CONCLUSION

We have developed a self-consistent algorithm, in accordance with the principle of maximum entropy, that calculates the chemical potentials that produce experimentally observed mean particle numbers. With this method and the grand canonical Monte Carlo simulations, we have gained insight into the mechanism and underlying principles of hierarchical assembly of macromolecular complexes, with emergent features varying with the crowding content.

Our method is a framework to connect the growing proteomic (or other "-omic") information(21, 59) to physical models. We attain the protein abundance from PaxDb;(41) however, any other methods for extracting the experimentally observed relative abundance can be used. In order to establish this new method of finding the correct chemical potentials given the cell type, we used the simplest protein-protein interactions in our structure-based Hamiltonian (section II.D) that contributes to the quinary interactions in the formation of higher-order protein complexes. Many studies have focused on understanding the protein-protein interaction networks(32, 33, 60), and our study here brings to attention the importance of the chemical potential for protein-complex assembly in the spatial arrangement of proteins in quinary complexes. Since the state of the cell may change both the interaction between proteins and chemical potentials, understanding the relationship between both of these aspects will be an important future work. Our method lays the foundation to create physical models that are bioinformatically consistent.

**ACKNOWLEDGMENTS**

We thank the computing resources from the Hewlett Packard Enterprise Data Science Institute at UH. A.G.G. was supported by a training fellowship on the Houston Area Molecular Biophysics Program (T32 GM008280). We also thank funding from the National Science Foundation (MCB-1412532, OAC-1531814).

**Appendix A: Single-Component Fluid Interaction Strength Analysis**

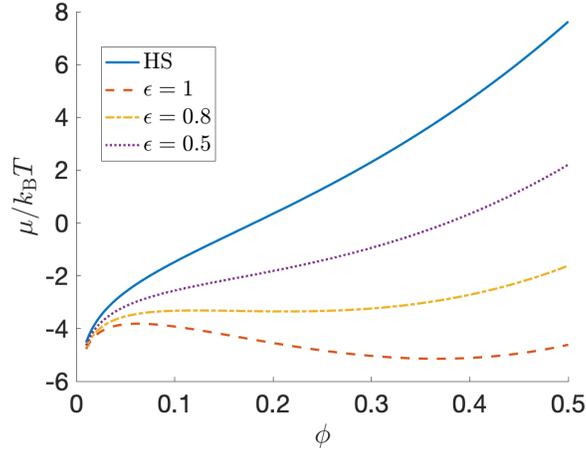

**FIG 9**. Chemical potential as a function of total volume fraction (ϕ) of a single-component fluid with LJ interactions at various interaction strengths. HS is hard spheres.

We chose $\epsilon = 0.5$ for the Lennard-Jones potential interaction strength, because above 0.8, certain values of the chemical potential corresponded to multiple volume fractions, which signify multiple phases (shown in appendix Fig A1.). Thus, a one-to-one correspondence of chemical potential to volume fraction reduces the chance of having large fluctuation in energy and/or density.

**Appendix B: Analytical Approximation of μ for Multi-Component Mixtures**

In Fig. 4, we compared the calculated chemical potential from our self-consistent algorithm with an analytical approximation of μ. Here we derive the equations used for those curves. In an ideal gas, $\mu = k_B T \ln\phi$. However, at non-dilute conditions such as the crowded cell,

$$\frac{\mu}{k_B T} = \ln\phi + \sum_{k=2}^{\infty} B^{(k)} \phi^{k-1},$$

where, $B^{(k)}$ is the $k^{th}$ virial coefficient. For multi-component mixture the second coefficient between particles $\alpha$ and $\beta$ can be calculated by,

$$B^{(2)}_{\alpha\beta} = -2\pi \int r_{\alpha\beta}^2 \left(e^{-\beta U_{LJ}(r_{\alpha\beta})} - 1\right) dr_{\alpha\beta}.$$

For higher order terms the integrals become increasingly more complex. Since the repulsive terms become more dominant as $\phi \to 1$, we can assume the potential of hard spheres and ignore the attractive terms.

$$\frac{\mu_\alpha}{k_B T} = \ln[\phi_\alpha(1 - \ln x_\alpha)] + \sum_j B^{(2)}_{\alpha\beta}\phi_\alpha + B^{(3)}_{HS}\phi^2 + B^{(4)}_{HS}\phi^3 + \mathcal{O}(\phi^4),$$

where $\phi_\alpha \equiv \phi \tilde{x}_\alpha^{exp}$, and $B^{(k)}_{HS}$ is the hard sphere virial coefficient found from the look-up table (61).